 \documentclass[preprint, preprintnumbers,amsmath,amssymb,amssymb]{revtex4}
\usepackage{graphicx}
\usepackage{dcolumn}
\usepackage{subfigure} 
\usepackage{bm}
\begin{document}
\title{Doubly charged Higgs in a 3-3-1 model at the CERN LHC\\
Prospects for Charged Higgs Discovery at Colliders \\
               16-19 September 2008\\
                   Uppsala, Sweden
}  
\author{J. E. Cieza Montalvo$^1$, Nelson V. Cortez$^{*,}$, M. D. Tonasse$^2$}   
\address{$^1$Rio de Janeiro State University, $^2$Sao Paulo State University, Brazil\\
$^{*,}$Speaker}  
\begin{abstract}
Doubly charged Higgs bosons are very important in particle physics nowadays because they can give mass to neutrinos through seesaw mechanisms. In this work we present the results in the searching for doubly charged Higgs at the CERN LHC for a 3-3-1 model. 
\end{abstract}
\maketitle
CERN LHC must bring great news in physics fields in the next years. In special, the discovery of Higgs boson and extra-dimensions occupy a very important part of the physicists' mind worldwide. Higgs bosons, for instance, can signalize to new models where these particles can be singly and$/$or doubly charged. There is a big interest in doubly charged Higgs bosons  (DCHBs) due to seesaw mechanisms which can give mass to neutrinos, as for instance in Ref. \cite{coton05}. \par
Very interesting results were obtained recently in the searching for DCHBs with lepton flavor violating decays involving tau leptons using Left-Right Symmetric model \cite{Aaltonen:2008ip}. However, an important class of models was raised through an efficient chiral expansion of the Standard Model(SM) in the electroweak sector, i.e. the full gauge symmetry passed from $SU(3)_C \otimes SU(2)_W \otimes U(1)_Y$ to $SU(3)_C \otimes SU(3)_L \otimes U(1)_N$, where $SU(2)_W \subset SU(3)_W$. For short it has been named 3-3-1 Model, see Ref. \cite{PPF92}. 
A very recent exciting manuscript has suggested that the observed excess of muons in studies in $p \bar{p}$ collisions at $\sqrt{s} = 1.96$ TeV arises from the decay of doubly-charged bosons in particles predicted in the 3-3-1 model \cite{Frampton:2008bf}.\par
The authors of this present work have worked with different kind of particle colliders searching for DCHBs as can be seen in Refs. \cite{cnt1a4}. They have worked with a 3-3-1 model which has the simplest scalar sector in this class of models \cite{PT93a}. An overview of it is presented below with results for the CERN LHC.\par
{\bf THE 3-3-1 MODEL:} The underlying eletroweak symmetry group is SU(3)$_L\otimes$U(1)$_N$, where N is the quantum number of the $U(1)$ group. Therefore, the left-handed lepton matter content is $\left(\begin{array}{ccc} \nu^\prime_a & \ell^\prime_a & {\tt L}^\prime_a\end{array}\right)^{\tt T}_L$ transforming as $\left({\bf 3}, 0\right)$, where $a = e, \mu, \tau$ is a family index (we are using primes for the interaction eigenstates). ${\tt L}^\prime_{aL}$ are lepton fields which can be the charge conjugates ${\ell^\prime_{aR}}^C$ \cite{PPF92} or the antineutrinos ${\nu^\prime_{La}}^C$ or heavy leptons $P^{\prime+}_{aL}$ $\left(P^{\prime+}_{aL} = E^{\prime+}_{L}, M^{\prime+}_{L}, T^{\prime+}_{L}\right)$ \cite{PT93a}. This model has the charge operator given by $\frac{Q}{e}$ = $\frac{1}{2} \left(\lambda_3 - \sqrt{3}\lambda_8 \right) + N$,
where $\lambda_3$ and $\lambda_8$ are the diagonal Gell-Mann matrices and $e$ is the elementary electric charge. The right-handed charged leptons are introduced in singlet representation of SU(3)$_L$ as $\ell^{\prime -}_{aR} \sim \left({\bf1}, -1\right)$ and $P^{\prime +}_{aR} \sim \left({\bf 1}, 1\right)$.\par
The model has three scalar triplets in the minimal scalar sector:
\begin{subequations}\begin{eqnarray}
\eta = \left(\begin{array}{c} \eta^0 \\ \eta_1^- \\ \eta_2^+ \end{array}\right) \sim \left({\bf 3}, 0\right), \quad \rho = \left(\begin{array}{c} \rho^+ \\ \rho^0 \\ \rho^{++} \end{array}\right) \sim \left({\bf 3}, 1\right), \quad 
\chi =
\left(\begin{array}{c} \chi^- \\ \chi^{--} \\ \chi^0 \end{array}\right) \sim \left({\bf 3}, -1\right).
\label{eigs}
\end{eqnarray}\label{eign1}\end{subequations}
The most general, gauge invariant and renormalizable Higgs potential, which conserves the leptobaryon number \cite{PT93b}, is
\begin{eqnarray}
V\left(\eta, \rho, \chi\right) & = & \mu_1^2\eta^\dagger\eta +
\mu_2^2\rho^\dagger\rho + \mu_3^2\chi^\dagger\chi +
\lambda_1\left(\eta^\dagger\eta\right)^2 +
\lambda_2\left(\rho^\dagger\rho\right)^2 +
\lambda_3\left(\chi^\dagger\chi\right)^2 + \cr && +
\left(\eta^\dagger\eta\right)
\left[\lambda_4\left(\rho^\dagger\rho\right) +
\lambda_5\left(\chi^\dagger\chi\right)\right] + \lambda_6
\left(\rho^\dagger\rho\right)\left(\chi^\dagger\chi\right) +
\lambda_7\left(\rho^\dagger\eta\right)\left(\eta^\dagger\rho\right)
+ \cr && +
\lambda_8\left(\chi^\dagger\eta\right)\left(\eta^\dagger\chi\right)
+
\lambda_9\left(\rho^\dagger\chi\right)\left(\chi^\dagger\rho\right)
+ \frac{1}{2}\left(f\epsilon^{ijk}\eta_i\rho_j\chi_k + {\mbox{H.
c.}}\right) \label{pot}\end{eqnarray} \ . \par
The neutral components of the scalars triplets (\ref{eigs}) develop non zero vacuum expectation values (VEV's) $\langle\eta^0\rangle = v_\eta$, $\langle\rho^0\rangle = v_\rho$ and
$\langle\chi^0\rangle = v_\chi$, with $v_\eta^2 + v_\rho^2 = v_W^2 =
(246 \mbox{ GeV})^2$. The pattern of symmetry breaking is $\mbox{SU(3)}_L\otimes\mbox{U(1)}_N\stackrel{\langle\chi\rangle}{\longmapsto} \mbox{SU(2)}_L\otimes\mbox{U(1)}_Y\stackrel{\langle\eta,
\rho\rangle}{\longmapsto}\mbox{U(1)}_{\rm em}$. Therefore, we can expect $v_\chi \gg v_\eta, v_\rho$. In the potential (\ref{pot}), $f$
and $\mu_j$ $\left(j = 1, 2, 3\right)$ are  constants with dimension of mass and the $\lambda_i$ $\left(i = 1, \dots, 9\right)$ are adimensional constants. The procedures to obtain the masses of the Higgs bosons are showed in Ref. \cite{TO96} under the conditions $v_\chi \approx -f$, leading a following  results for the masses of the neutral physical scalars
\begin{subequations}
\begin{equation}
m_{H_{1}^{0}}^{2} \approx 4\frac{\lambda_2v_\rho^4 - 2\lambda_1v_\eta^4}{v_\eta^2 - v_\rho^2}, \quad m_{H_{2}^{0}}^{2} \approx \frac{v_W^2v_\chi^2}{2v_\eta v_\rho},  \quad 
m_{H_{3}^{0}}^{2} \approx -\lambda_3v_\chi^2, \quad m_h^2 = -\frac{fv_\chi}{v_\eta v_\rho}\left[v_W^2 + \left(\frac{v_\eta v_\rho}{v_\chi}\right)^2\right] 
\label{m0} \ , \end{equation}
for the singly charged ones 
\begin{equation}
m_{\pm 1}^2 = \frac{v_W^2}{2v_\eta v_\rho}\left(fv_\chi - 2\lambda_7v_\eta v_\rho\right), \qquad 
m_{\pm 2}^2 = \frac{v_\eta^2 + v_\chi^2}{2v_\eta v_\chi}\left(fv_\rho - 2\lambda_8v_\eta v_\chi\right)
\label {m+} \ , \end{equation}
and for the doubly charged Higgs bosons 
\begin{equation}
m^2_{\pm \pm} = \frac{v_\rho^2 + v_\chi^2}{2v_\rho v_\chi}\left(fv_\eta - 2\lambda_9v_\rho v_\chi\right)
\label{m++} \ . \end{equation}
\end{subequations} \par
The total leptonic number is conserved \cite{PPF92} although that non standard field interactions violate leptonic number.
\begin{center}
{\bf RESULTS}
\end{center}

\begin{figure}
\begin{minipage}[b]{0.45\linewidth} 
\centering
\includegraphics[width=8.8cm]{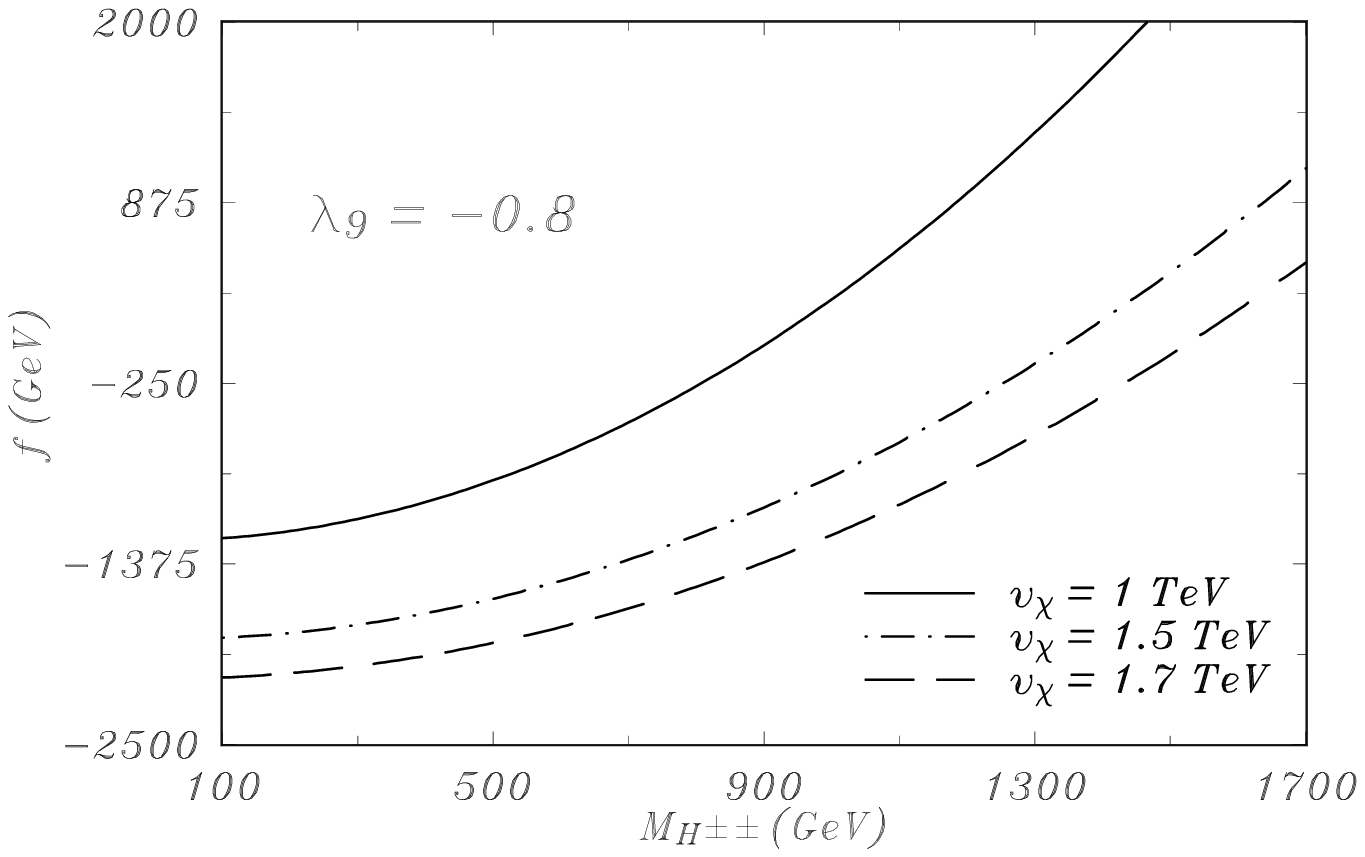}
\caption{\label{fig1} Parameter $f$ as a function of mass of doubly charged Higgs boson for the parameter $\lambda_{9}=$ -0.8.}
\end{minipage}
\hspace{0.5cm} 
\begin{minipage}[b]{0.45\linewidth}
\centering
\includegraphics[width=8.8cm]{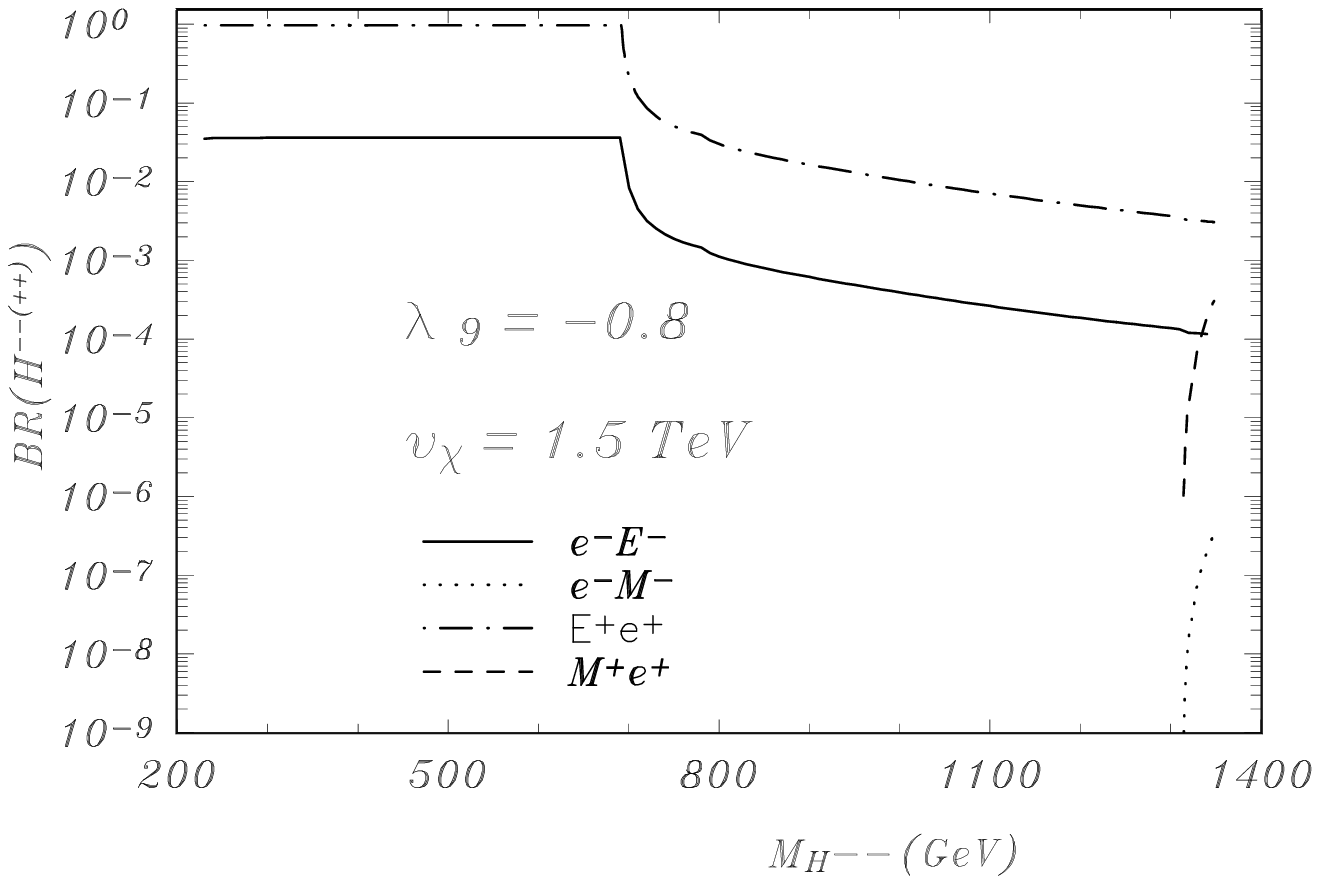}
\caption{\label{fig3}  Branching ratios for the doubly charged Higgs decays as a functions of $m_{H^{\pm \pm}}$ for $\lambda_{9}=$ -0.8 for the fermionic sector.}
\end{minipage}
\end{figure}

\begin{figure}
\begin{minipage}[b]{0.45\linewidth} 
\centering
\includegraphics[width=8.8cm]{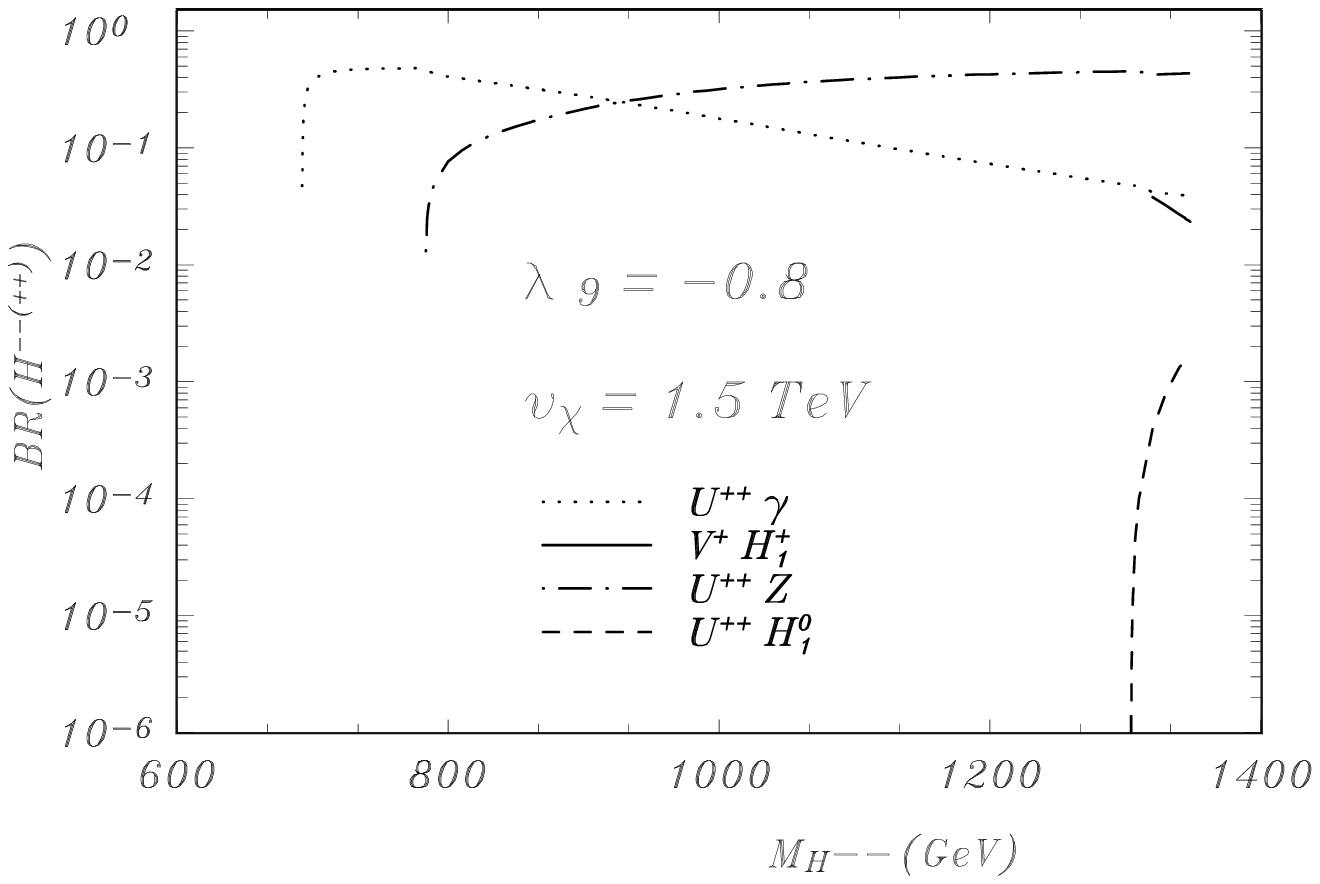}
\caption{\label{fig4} Branching ratios for the doubly charged Higgs decays as a functions of $m_{H^{\pm \pm}}$ for $\lambda_{9}=$ -0.8 for the bosonic  sector.}
\end{minipage}
\hspace{0.5cm} 
\begin{minipage}[b]{0.45\linewidth}
\centering
\includegraphics[width=8.8cm]{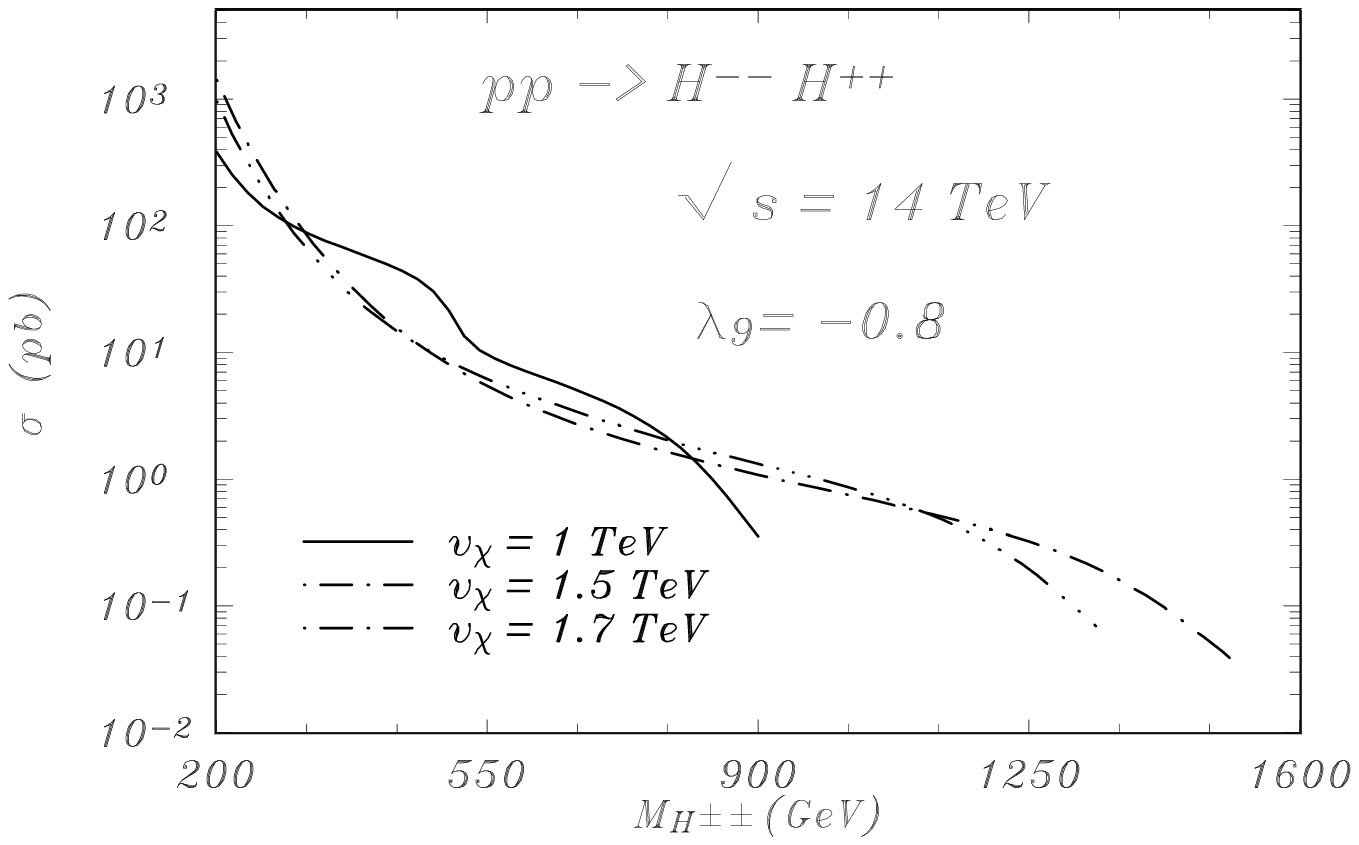}
\caption{\label{fig8}  Total cross section for the process $p p \rightarrow H^{--} H^{++}$ as a function of $m_{H^{\pm \pm}}$ for $\lambda_{9}=-0.8$}
\end{minipage}
\end{figure}
When we consider particular values for the adimensional parameter $\lambda_9$ in the Higgs potential (\ref{pot}) and for the vacuum expectation value $v_\chi$ \cite{TO96} and with all constraints in the model used in this work, so the production  and signals for the DCHBs in the 3-3-1 models at LHC collider are very significant.
We have fixed the general bounds of the adimensional constants of the Higgs potential (\ref{pot}) as $-3 \leq \lambda_i \leq 3$ $\left(i = 1, \ldots, 9\right)$ to guarantee approximately the perturbative regime. When we consider $-f \approx v_\chi \gg v_\eta, v_\rho$, we have
\begin{equation}
\lambda_4 \approx 2\frac{\lambda_2v_\rho^2 - \lambda_1v_\eta^2}{v_\eta^2 - v_\rho^2}, \qquad \lambda_5v_\eta^2 + 2\lambda_6v_\rho^2 \approx -\frac{v_\eta v_\rho}{2}
\label{vin1}
\end{equation}
as seen in Ref. \cite{TO96}. \par
A analysis of Eq. (\ref{m++}) also shows that the lowest values for $\lambda_{9}$ and $f$, such to produce values for $m_{H^{\pm \pm}}$ between 100 GeV and 200 GeV, for $v_\chi$ = 1500 GeV and  $v_\rho$ = 195 GeV are $\lambda_{9}= -4.50 \times  10^{-3}$, $f= -0.23 $ GeV and $\lambda_{9}= -1.80 \times 10^{-2}$, $f= -0.92$ GeV. The values  for the parameters $\lambda_{5}$ and $\lambda_{6}$ lead  to the constraint $v_{\eta}>40.5$ GeV, which can be seen as following from the Eq. (\ref{vin1}). In the approximation $-f \approx v_{\chi}$ it is appropriate to choose the parameter  $-1.2 \leq \lambda_{9} \leq -0.8$ for the mass of $m_{H^{\pm \pm}}=500$ GeV. For simplicity we present here only the results for $\lambda_{9}= -0.8$. The VEV's values $v_\eta$ and $v_\rho$ influence very little on the mass of $m_{H^{\pm\pm}}$, this happen because in the Eq.  (\ref{m++}) the $v_{\eta}$ and $v_{\rho}$ are related by the relation $v_{\eta}^{2}+ v_{\rho}^{2}= (246 \mbox{ GeV})^{2}$  and $v_{\rho}$ is smaller compared with $v_{\chi}$. \par  
We have chosen for the parameters, masses and the VEV, the following representative values: $\lambda_{1} =-1.2$,  $\lambda_{2}=\lambda_{3}=-\lambda_{6}=\lambda_{8}=-1$, $\lambda_{4}= 2.98$ $\lambda_{5}=-1.57$, $\lambda_{7}=-2$ and $\lambda_{9}=-0.8$, $v_{\eta}=195$ GeV and $v_{\chi}=1500$ GeV. Others particles, such as $J_{1,2,3}, T, U^{\pm \pm}, Z', H_{2}^{\pm}, H_{2,3}^{0}$ and $h^{0}$, do not take part in this process because of their masses are larger than the mass of the DCHBs. 
Observing the Fig. 1 we see that for the $v_{\chi}=1000$ GeV we have the acceptable masses up to $m_{H^{\pm \pm}} \simeq 903$ GeV, for $v_{\chi}=1500$ GeV we will have up to $m_{H^{\pm \pm}} \simeq 1346$ GeV and for $v_{\chi}=1700$ GeV we have consequently up to $m_{H^{\pm \pm}}\simeq 1525$ GeV. Figs. \ref{fig3} and \ref{fig4}, show the branching ratios for the Higgs decays, $H^{\pm \pm} \rightarrow {\it all}$. 
We have chosen $v_{\chi}=1700$ GeV as the maximum value because it is constraint by the mass of the  $m_{Z'}=(0.5-3)$ TeV, which is proportional to $v_{\chi}$ \cite{PPF92}\par
Considering that the expected integrated luminosity for the LHC will be of order of $3 \times 10^5$ pb$^{-1}$/yr then the statistics give a total of $\simeq 3.9 \times 10^{6}$ events per year for Drell-Yan process, if we take the mass of the Higgs boson $m_{H^{\pm \pm}}= 500$ GeV, $v_{\chi}=1500$ GeV, $m_{H_{2}}^{\pm}= 1163.7$ GeV, $m_h = 2229.9$ GeV and $\lambda_{9}$ = $-0.8$. Considering that the signal for $H^{\pm \pm}$ production will be $e^{-} P^{-}$ and $e^{+} P^{+}$ and taking into account that the branching ratios for both particles we would have approximately $1.3 \times 10^{5}$ events per year for Drell-Yan process. Taking now the mass of the Higgs boson $m_{H^{\pm \pm}}= 700$ GeV,  $v_{\chi}=1500$ GeV, $m_{H_2^\pm}= 1223.6$ GeV, $m_h = 2052.2$ GeV and the same value of $\lambda_{9}$ as above. We then have a total of $\simeq 2.2 \times 10^6$ events per year for Drell-Yan process. 
Fig. \ref{fig8} shows the cross section for the process $pp \to H^{++}H^{--}$.
\begin{center}
\begin{table}[t]
\caption{\label{tab2}\footnotesize\baselineskip = 12pt Total number of events per year for DCHBs and with respect the branching ratios from $H^{\pm\pm} \to e^{+}P^{+} + e^{-}P^{-}$ in function of $m_H^{++}$, $m^{\pm}_{H_2}$, $m_h$ and $v_\chi$, which their units are in GeV.}
\begin{tabular}{|c|c|c|c|c|c|}
\hline\hline
$v_\chi$ & $m_H^{++}$ & $m^{\pm}_{H_2}$ & $m_h$ & Total events per year & Events per year for BR($H^{\pm\pm} \to e^{+}P^{+} + e^{-}P^{-}$) \\
\hline \hline
\raisebox{-1.5ex}{$0.8$} &
$500$ & 1163.7 & 2222.9 & 39 $\times$ $10^6$ & 1.3 $\times$ $10^5$\\
& $700$ & 1223.6 & 2052.2 & 2.2 $\times$ $10^6$ & 3.9 $\times$ $10^{3}$ \\
\hline \hline
\end{tabular}
\end{table}
\end{center}
The main background for this signal, $p \ p \rightarrow H^{--} H^{++}  \rightarrow e^{-} P^{-} \ (e^{+} P^{+} )$, could come from the process $p \ p \rightarrow Z \ Z$ and another small background from the $p \ p \rightarrow W^{-} W^{+} Z$, all these backgrounds can be eliminated (see Ref. \cite{cp02}).
Our studies have indicated the possibility of obtaining a clear signal of the doubly charged Higgs bosons, with a satisfactory number of events, in the next particle colliders and in a very special way to the CERN LHC, as can be seen in Table I.
 
\end{document}